\documentclass[10pt,a4paper,twocolumn,fleqn]{article}
\usepackage{latexsym}
\usepackage{graphicx}
\usepackage{times}

\title{\textbf{ An Identity Based Key Management Scheme in Wireless Sensor Networks}}
\author{Ashok Kumar Das\\
Department of Computer Science and Engineering\\
Indian Institute of Technology, Kharagpur 721 302, India\\
\tt\char`akdas\char`@cse.iitkgp.ernet.in\\ \\
Debasis Giri\\
Department of Mathematics\\
Indian Institute of Technology, Kharagpur 721 302, India\\
\tt dgiri@maths.iitkgp.ernet.in\\
}
\date{}

\begin{document}
\maketitle
\thispagestyle{empty}
\begin{abstract}
\noindent \emph{Pairwise key establishment is one of the fundamental security services in sensor networks which enables sensor nodes in a sensor network to communicate securely with each other using cryptographic techniques. It is not feasible to apply traditional public key management techniques in resource-constrained sensor nodes, and also because the sensor nodes are vulnerable to physical capture. In this paper, we introduce a new scheme called the identity based key pre-distribution using a pseudo random function (IBPRF), which has better trade-off between communication overhead, network connectivity and resilience against node capture compared to the other key pre-distribution schemes. Our scheme can be easily  adapted in mobile sensor networks. This scheme supports the addition of new sensor nodes after the initial deployment and also works for any deployment topology. In addition, we propose an improved version of our scheme to support large sensor networks.}
\end{abstract}

\section{Introduction}

 In a sensor network, many tiny computing nodes called sensors, are scattered in an area for the purpose of sensing some data and transmitting the data to nearby \emph{base stations} for further processing. The transmission between the sensors is done by short range radio communications. The base station is assumed to be computationally well-equipped whereas the sensor nodes are resource-starved. Such networks are used in many applications including tracking of objects in an enemy's area for military purposes, distributed seismic measurements, pollution tracking, monitoring fire and nuclear power plants, tacking patients, engineering and medical explorations like wildlife monitoring, etc. Mostly for military purposes, data collected by sensor nodes need be encrypted before transmitting to neighboring nodes and base stations. \\

\indent The following issues make secure communication between sensor networks different from usual (traditional) networks:
\begin{itemize}
\item \emph{Limited resources in sensor nodes:} Each sensor node contains a primitive processor featuring very low computing speed and only small amount of programmable memory. An example is the popular Atmel ATmega 128L processor.
\item \emph{Limited life-time of sensor nodes:} Each sensor node is battery-powered and is expected to operate for only few days. Therefore, once the deployed sensor nodes expire, it is necessary to add some fresh nodes for continuing the data collection operation. This is referred to as the \emph{dynamic management of security objects (like keys)}.
\item \emph{Limited communication abilities of sensor nodes:} Sensor nodes have the ability to communicate each other and the base stations by the short range wireless radio transmission at low bandwidth and over small communication ranges (typical example is 30 meters (100 feet)).
\item \emph{Lack of knowledge about deployment configuration:} Most of cases, the post deployment network configuration is not known a priori. As a result, it is unreasonable to use security algorithms that have strong dependence on locations of sensor nodes in a sensor network.
\item \emph{Mobility of sensor nodes:} Sensor nodes may be mobile or static. If sensor nodes are mobile then they can change the network configuration at any time. 
\item \emph{Issue of node capture:} A part of the network may be captured by the adversary/enemy. The resilience measurement against node capture is computed by comparing the number of nodes captured, with the fraction of total network communications that are exposed to the adversary \emph{not including} the communications in which the compromised nodes are directly involved.
\end{itemize}

\indent Thus, it is not feasible to use public-key cryptosystems in resource constrained sensor networks. Hence, only the symmetric cipher such as DES/IDEA/RC5~\cite{bk:02,jr:04} is the viable option for encryption/decryption of secret data. But setting up symmetric keys among communication nodes is a challenging task in a sensor network. A survey on sensor networks can be found in~\cite{jr:06,jr:09}. \\

\indent The topology of sensor networks changes due to the following three phases:
\begin{itemize}
\item \textit{Pre-deployment and deployment phase:} Sensor nodes can be deployed from the truck or the plane in the sensor field.
\item \textit{Post-deployment phase:} Topology can change after deployment because of irregularities in the sensor field like obstacles or due to jamming, noise, available energy of the nodes, malfunctioning, etc., or due to the mobile sensor nodes in the network.
\item \textit{Redeployment of additional nodes phase:} Additional sensor nodes can be redeployed at any time to replace the faulty or compromised sensor nodes.
\end{itemize}

\indent A protocol that establishes cryptographically secure communication links among the sensor nodes is
called the \emph{bootstrapping protocol}. Several methods~\cite{cf:01,cf:02,cf:04,cf:07} are already proposed
in order to solve the bootstrapping problem. All these techniques are based on random deployment models, that
is, they do not use the pre-deployment knowledge of the deployed sensor nodes. Eschenauer and Gligor~\cite{cf:01} proposed the basic random key predistribution called the EG scheme, in which each sensor is assigned
a set of keys randomly selected from a big key pool of the keys of the sensor nodes. Chan et al.~\cite{cf:02} 
proposed the $q$-composite key predistribution and the random pairwise keys schemes. For both the EG
and the $q$-composite schemes, if a small number of sensors are compromised, it may reveal to compromise a large fraction of pairwise keys shared between non-compromised sensors. However, the random pairwise keys predistribution is perfectly secure against node captures, but there is a problem in supporting the large network. Liu and Ning's polynomial-pool based key predistribution scheme~\cite{cf:04} and the matrix-based key predistribution 
proposed by Du et al.~\cite{cf:05} improve security considerably. Liu and Ning~\cite{jr:10} proposed the extended version of the closest pairwise keys scheme~\cite{cf:09} for static sensor networks. Their scheme is based on the pre-deployment locations of the deployed sensor nodes and a pseudo random function (PRF) proposed by Goldreich et al.~\cite{jr:08}. There is no communication overhead for establishing direct pairwise keys between neighbor nodes and the scheme is perfectly secure against node capture. \\

\indent The rest of the paper is organized as follows. Section 2 describes our proposed scheme called the identity based key predistribution using a pseudo random function (IBPRF). In Section 3, we provide a theoretical analysis for this scheme. In Section 4, we discuss the security issues with respect to our scheme. In Section 5, we provide an improved version of our scheme for distributed sensor networks. In Section 6, we compare our scheme with the previous schemes~\cite{cf:01,cf:02,cf:04} with respect to communication overhead, network connectivity, and resilience against node captures. Finally, Section 7 concludes the paper.

\section{Identity Based Key Pre-Distribution using a Pseudo Random Function (IBPRF)}

The bootstrapping protocol for the random key predistribution schemes ~\cite{cf:01,cf:02,cf:04} incurs much 
more communication overhead for establishing direct pairwise keys between sensor nodes in a sensor 
network. Our goal is to design a protocol which basically reduces the communication overhead for establishing
direct pairwise keys between sensors during direct key establishment phase of the bootstrapping. We propose a
 new scheme called the \emph{identity based key predistribution using a pseudo random function} (IBPRF), which 
serves our above desired purpose. \\

\indent IBPRF has the following interesting properties:
\begin{itemize}
\item There is no communication overhead during direct key establishment phase for establishing direct pairwise keys between sensors.
\item There is no communication overhead during the addition of new sensor nodes.
\item When the sensor nodes are mobile, our scheme easily establish direct pairwise keys between the mobile sensor nodes and their physical neighbors with which they do not share keys currently with some desired
probability.
\item It works for any deployment topology.
\end{itemize}

\indent IBPRF is based on the following two ingredients:
\begin{itemize}
\item A pseudo random function (PRF) proposed by Goldreich et al. in 1986~\cite{jr:08}.
\item A master key (MK) shared between each sensor node and the key setup server.
\end{itemize}

\indent The different phases for this scheme are as follows. \\
\subsection{Key Pre-Distribution}
Let $N$ be a pool of the ids of $n$ sensor nodes in a sensor network. Assume that each sensor node $u$ is capable of holding a total of $m+1$ cryptographic keys in its key ring $K_{u}$. The key predistribution has the following steps:
\begin{itemize}
\item \emph{Step-1:} For each sensor node $u$, the key setup server randomly generates a master-key 
$MK_{u}$.
\item \emph{Step-2:} For each sensor node $u$, the key setup server selects a set $S$ of $m$ randomly 
generated ids of sensor nodes from the pool $N$ which are considered as the probable physical neighbors'
ids. Let $S$ = \{$v_{1},v_{2}, \ldots ,v_{m}$\}. For each node id $v_{i}\in S$ $(i=1,2, \ldots ,m)$, the
key setup server generates a symmetric key $SK_{u,v_{i}} = \mathit{PRF}_{MK_{v_{i}}}(u)$ as the pairwise key
shared between the nodes $u$ and $v_{i}$, where $MK_{v_{i}}$ is the master key for $v_{i}$ and $u$ is the id
of the node $u$.
\end{itemize}

\indent  For each $v_{i}\!\in\! S$, the key-plus-id combination $(SK_{u,v_{i}},v_{i})$ is stored in $u$'s key ring $K_{u}$.
 We note that each node $v_{i}$ can easily compute the same key $SK_{u,v_{i}}$ with its master
key and the id of node $u$. The sensor node  $v$ is called a master sensor node of $u$ if the shared key
 between them is calculated by $SK_{u,v} = \mathit{PRF}_{MK_{v}}(u)$. In other words, node $u$ is called a
 slave sensor node of $v$ if $v$ is a master sensor node of $u$.

\subsection{Direct Key Establishment}
After deployment of sensor nodes in a deployment area (i.e., target field), sensor nodes will establish direct pairwise keys
between them. Direct key establishment phase has the following steps:
\begin{itemize}
\item \emph{Step-1:} Each sensor node first locates its all physical neighbors. Nodes $u$ and $v$ are called \emph{physical neighbors} if they are within the communication range of one another. They are called \emph{key neighbors} if they share a pairwise key. They are said to be \emph{direct neighbors} if they are both physical as
well as key neighbors. Now, after identifying the physical neighbors by a sensor node $u$, it can easily verify which ids of the physical neighbors exist in its key ring $K_{u}$. If $u$ finds that it has the predistributed pairwise key $SK_{u,v} = \mathit{PRF}_{MK_{v}}(u)$ with node $v$ then it informs sensor $v$ that it has such a key. This notification is done by sending a short message containing the id of node $u$ that $u$ has such a key. We note that this message never contains the exact value of the key $SK_{u,v}$.
\item \emph{Step-2:} Upon receiving such a message by node $v$, it can easily calculate the shared 
pairwise key $SK_{u,v} = \mathit{PRF}_{MK_{v}}(u)$ by using its own master key and the id of node $u$.
\end{itemize}

\noindent Thus, nodes $u$ and $v$ can establish a direct pairwise key shared between them very easily and use this key for their future communication.

\subsection{Path Key Establishment}
This is an optional stage, if requires, adds the connectivity of the network. After direct key establishment,
if the connectivity is still poor, nodes $u$ and $v$ which are physical neighbors not sharing a pairwise key, can establish a direct key between them as follows.
\begin{itemize}
\item \emph{Step-1:} $u$ first finds a path $\langle u=u_{0},u_{1},u_{2}, \ldots ,u_{h-1},u_{h}=v \rangle$ such that
each $(u_{i},u_{i+1})$ $(i=0,1,2, \ldots ,h-1)$ is a secure link.
\item \emph{Step-2: } $u$ generates a random number $k'$ as the shared pairwise key between $u$ and $v$ and encrypts it using the shared key $SK_{u,u_{1}}$ and sends to node $u_{1}$.
\item \emph{Step-3:} $u_{1}$ retrieves $k'$ by decrypting the encrypted key using $SK_{u,u_{1}}$ and
encrypts it using the shared key $SK_{u_{1},u_{2}}$ between $u_{1}$ and $u_{2}$ and sends to $u_{2}$.
\item \emph{Step-4:} This process is continued until the key $k'$ reaches to the desired destination
node $v$.
\end{itemize}
\noindent As a result, nodes $u$ and $v$ use $k'$ as the direct pairwise key shared between them for future communication. Since this process involves more communication overhead to establish a pairwise key between nodes, in practice $h=2$ or $3$ is recommended.

\subsection{Mobility of Sensor Nodes}
Suppose that a sensor node $u$ moves from one location to another. Due to location updation of $u$, the
connectivity of $u$ with the new neighbors may also change. In the new location, assume that $u$ finds the
ids of its some new physical neighbors with which it does not currently share any keys. If $v$ be one such physical neighbor, $u$ informs to $v$ that it has a pairwise key with $v$. This notification takes place by sending a request message to $v$ containing the id of sensor node $u$ excluding the exact value of the key. Upon receiving this message, $v$ can immediately compute the pairwise key shared between them by executing one efficient PRF operation and by using the master key $MK_{v}$ for $v$ and the id of sensor node $u$. Thus, $u$ and $v$ use this key for their future communication. \\

\indent After performing this stage, if sensor node $u$ finds still poor connectivity, it may opt for at most 1-hop path key establishment because path key establishment involves more communication overhead. Of course we assume that mobility of sensor nodes are infrequent.

\subsection{Addition of Sensor Nodes}
In order to add a new sensor node $u$, the key setup server selects a set $S$ of $m$ randomly generated ids
of sensor nodes from the pool $N$. The key setup server randomly generates a master key $MK_{u}$ for node $u$.  For each sensor node id $v \in S$, the key setup server takes the master key $MK_{v}$ and compute the secret key $SK_{u,v} = \mathit{PRF}_{MK_{v}}(u)$ as the shared pairwise key between nodes $u$ and $v$, and distributes the key-plus-id combination $(SK_{u,v},v)$ to $u$. After deployment of sensor node $u$, it establishes direct pairwise keys using direct key establishment phase of IBPRF with the physical neighbors for which the ids are in $u$'s key ring $K_{u}$. \\

\indent Now, if $u$ finds still poor connectivity after direct key establishment, it can perform path key establishment stage with 2 or 3 hops.

\section{Analysis}
In this section, we shall now compute the probability of establishing direct keys between two sensor nodes during direct key establishment, and the probability of establishing a pairwise key between two sensor nodes during path key establishment. We shall also analyze the storage overhead and the communication overhead required by our scheme.

\subsection{Probability of Establishing Direct Keys}
Let $p$ be the probability that two physical neighbors can establish a direct pairwise key. For the derivation
of $p$, we first observe that two physical neighbors $u$ and $v$ can establish a pairwise key only if the
key ring $K_{u}$ of node $u$ contains the shared secret key $SK_{u,v} = \mathit{PRF}_{MK_{v}}(u)$ and the id
 of node $v$, or the key ring $K_{v}$ of node $v$ contains the shared secret key $SK_{v,u} = \mathit{PRF}_{MK_{u}}(v)$ and the id of node $u$ because of the fact that any of nodes $u$ and $v$ can initiate for establishing a pairwise key between them. \\

\indent We then have, $p = 1 -$ (probability that both $u$ and $v$ do not establish a pairwise key). The total number of ways to select $m$ ids from the pool $N$ of size $n$ is $ \left( \begin{array}{c} n \\ m \end{array} \right) $. For a fixed key ring $K_{u}$ of node $u$, the total number of ways to select $K_{v}$ of a node $v$
such that $K_{v}$ does not have the id of $u$ is $ \left( \begin{array}{c} n-1 \\ m \end{array} \right) $.  Thus, we have
\begin{equation}
p = 1 - \frac{\left( \begin{array}{c}
                 n-1 \\ m
                 \end{array} \right) }
        {\left( \begin{array}{c}
                 n \\ m
                 \end{array} \right) } 
= \frac{m}{n}.
\end{equation}
  
\noindent We note that $p$ strictly depends on the network size $n$ and the key ring size. \\

\begin{figure}[h]
\centering
\includegraphics[scale=0.31,angle=270]{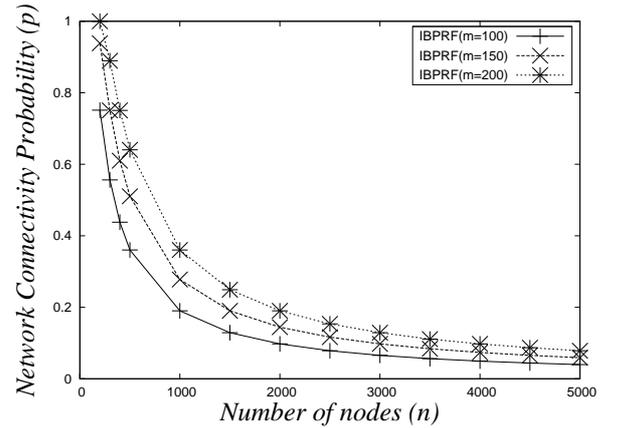}
\caption{The probability $p$ that two sensors establish a direct pairwise key v.s. the network size 
$n$, with $m = 100,150,200$.} \label{Figure: }
\end{figure}
\begin{figure}[h]
\centering
\includegraphics[scale=0.31,angle=270]{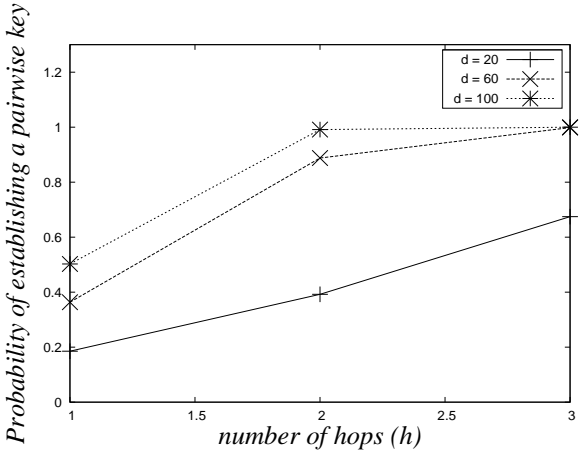}
\caption{The probability $p_{s}$ of establishing a pairwise key v.s. the probability $p$ that two sensor nodes establish a direct pairwise key, with $d = 20,60,100$.} \label{Figure: } 
\end{figure}

\noindent It is clear from Figure 1 that when the network size is small, our scheme provides better connectivity. Therefore, our scheme can not support a large network. In section 5, we have proposed an improved version of our scheme to support large networks.

\subsection{Probability of Establishing Keys using 1-hop Path Key Establishment}
If $d$ be the average number of neighbor nodes that each sensor node can contact, it follows from the similar
analysis in~\cite{cf:04} that the probability of two sensor nodes establishing a pairwise key (directly or
indirectly) is 
\begin{equation}
p_{s} = 1 - (1-p)(1-p^{2})^{d}.
\end{equation}

\indent The network connectivity probabilities for path key establishment with 1-hop are plotted in Figure 2. From this figure it is also clear that we are able to achieve better connectivity after executing this stage even if the network is almost disconnected initially.

\subsection{Calculation of Storage Overhead}
Each sensor node has to store a master key which is shared with the key setup server and $m$ predistributed
key-plus-id combinations. Hence, our scheme requires a storage overhead of maximum $m+1$ keys for each sensor node.

\subsection{Calculation of Communication Overhead}
For establishing a pairwise key between two sensor nodes $u$ and $v$, one of them, say $u$, initiates a request message to node $v$ that its key ring contains the shared key between them. Then, after receiving such a request,
node $v$ computes the shared key between $u$ and $v$ by performing only one efficient PRF operation. 
Hence, the communication overhead involves only one short message for informing the other node that it has a
pairwise key and the computational overhead due to single efficient PRF operation.

\section{Security Considerations}
The security of IBPRF depends on the following facts:
\begin{itemize}
\item The security of PRF~\cite{jr:08}.
\item A node's master key MK which is shared with the key setup server.
\end{itemize}

\indent It is observed that if a node's master key is not disclosed, no matter how many pairwise keys
generated by this master key are disclosed, the task is still computationally difficult for an adversary to
recover the master key MK as well as the non-disclosed pairwise keys generated with different ids of sensor
nodes. Again, each pre-distributed pairwise key between two sensor nodes is generated by using PRF function randomly. Thus, no matter how many sensor nodes are compromised, the direct pairwise keys between non-compromised nodes are still secure. In other word, node compromise does not eventually lead to compromise of the direct pairwise keys between the other non-compromised nodes. In this way, our scheme provides perfect security against node captures.

\section{ The Improved Scheme}
We note that our basic scheme (IBPRF) provides better connectivity if the network size is small, whereas it provides perfect security against node captures. In fact, there is no communication overhead during establishment of the direct pairwise keys between sensors and also during the addition of nodes after their initial deployment. \\

\indent To support a large sensor network, we wish to apply our basic scheme in distributed sensor networks. The deployment region is divided into $c$ number of sub-regions called the cells such that each cell can communicate with the base stations comfortably. Let the $i$-th cell be denoted by $\mathit{cell}_{i}$. Assume that each $\mathit{cell}_{i}$ contains $n_{i}$ number of sensor nodes. In practical situation, it is not always possible to deploy each node to a pre-determined location in the deployment region. We further assume that the key setup server only knows the nodes containing to a particular cell which will be deployed in that region randomly. In practice, this assumption is appropriate. Under this configuration, we now apply our basic scheme to each cell as follows. \\

\indent Let $N_{i}$ be the pool of the ids of $n_{i}$ sensor nodes in a cell $\mathit{cell}_{i}$. Assume that each sensor node $u$ is capable of holding a total of $m+1$ cryptographic keys. In key pre-distribution phase, for each node $u\in \mathit{cell}_{i}$, the key setup server randomly generates a master key $MK_{u}$. For each node $u\in \mathit{cell}_{i}$, the key setup server also selects a set $S$ of $m$ randomly generated ids of the sensor nodes from the pool $N_{i}$. For each $v\in S$, the key setup server generates a symmetric key $SK_{u,v} = \mathit{PRF}_{MK_{v}}(u)$ as the pairwise key shared between nodes $u$ and $v$, where $MK_{v}$ is the master key for node $v$ and $u$ is the id for node $u$. The key-plus-id combination $(SK_{u,v},v)$ is stored in $u$'s key ring $K_{u}$. After deployment of the sensor nodes, they establish direct pairwise keys using direct key establishment phase of our basic scheme (IBPRF). The other phases like path key establishment, mobility of sensor nodes, and addition of sensor nodes remain same as our basic scheme. \\

\indent Thus, sensor nodes in each cell establish pairwise keys between them and communicate with each other in that cell securely. For mobility of the sensor nodes, we restrict the sensor nodes to move in a particular cell only. \\

\indent Let $p_{i}$ denote the probability that two sensor nodes in the $i$-th cell $\mathit{cell}_{i}$ can establish a direct pairwise key between them. Similar to analysis in 3.1, we have
\begin{equation}
p_{i} = 1 - \frac{\left( \begin{array}{c}
                 n_{i}-1 \\ m
                 \end{array} \right) }
        {\left( \begin{array}{c}
                 n_{i} \\ m
                 \end{array} \right) } 
= \frac{m}{n_{i}}.
\end{equation}
     
\noindent The (average) probability that two sensor nodes in a network of size $n=\sum_{i=1}^{c} n_{i}$, establish a direct pairwise key between them is given by
\begin{equation}
p = \frac{\sum_{i=1}^{c} p_{i}}{c}.
\end{equation}

\indent Hence, we are able to achieve better connectivity for the entire network by using our improved version and selecting the appropriate size of the cells. However, the communication overhead as well as resilience measurement against node captures remain same as our basic scheme (IBPRF). We note that this improved scheme may not always work for ad hoc mode sensor networks.
 
\section{Comparison with Previous Schemes}
In this section, we compare both our basic scheme (IBPRF) and the improved scheme with the EG~\cite{cf:01}, the $q$-composite (qC)~\cite{cf:02}, and the polynomial-pool based~\cite{cf:04} schemes with respect to the communication overhead, network connectivity and resilience against node captures. \\

\noindent \emph{(1) Communication overhead} \\
\indent For the EG and the $q$-composite schemes, when a node wishes to establish pairwise keys with its
physical neighbor nodes, it needs to send a list of some messages encrypted by keys in its key ring. \\

\indent In case of the polynomial-pool based scheme, a sensor node also needs to send a list of some messages
encrypted by potential pairwise keys based on its polynomial shares for establishing a direct pairwise key
with a physical neighbor. \\

\indent Thus, the communication overhead is on the order of the key ring size for these schemes. But, for our schemes, the communication overhead is only due to one short message sent by a node to inform its physical neighbor that it has a pairwise key in its key ring and a single efficient PRF operation for computing the shared key $SK$ by the physical neighbor. Hence, both our basic scheme (IBPRF) and the improved scheme have much less communication overhead than the EG, the $q$-composite, and the polynomial-pool based schemes. \\  

\noindent \emph{(2) Resilience against node capture} \\  
\indent From the analysis of the EG scheme~\cite{cf:01} and the $q$-composite scheme~\cite{cf:02}, it follows that even if the number of nodes captured is small, these schemes may reveal a large fraction of pairwise keys shared between non-compromised sensors. The analysis of the polynomial-pool based scheme~\cite{cf:04} shows that this scheme is unconditionally secure and $t$-collusion resistant. Thus, it has better resilience against node captures than the EG and the $q$-composite schemes. However, both our basic scheme (IBPRF) and the improved scheme provide perfect security against node captures. \\

\noindent \emph{(3) Network connectivity} \\
\indent For the EG scheme~\cite{cf:01}, the probability of establishing a direct pairwise key between two sensor nodes is 
\begin{equation}
p_{EG} = 1 - \frac{\left( \begin{array}{c}
                 M-m \\ m
                 \end{array} \right) }
        {\left( \begin{array}{c}
                 M \\ m
                 \end{array} \right) } 
=1 - \prod_{i=0}^{m-1} \frac{M-m-i}{M-i}
\end{equation}
 
\noindent where $M$ and $m$ are the key pool size and key ring size of a sensor node. \\

\indent For the $q$-composite scheme~\cite{cf:02}, the probability of establishing a direct pairwise key between two
sensor nodes is
\begin{equation}
p_{qC} = 1 - \sum_{i=0}^{q-1} p_{i}
\end{equation}

\noindent where $p_{i} = \frac{{\left( \begin{array}{c}
                 M \\ i
                 \end{array} \right)}
                  {\left( \begin{array}{c}
                 M-i \\ 2(m-i)
                 \end{array} \right)}
                 {\left( \begin{array}{c}
                 2(m-i) \\ m-i
                 \end{array} \right)} }
        {\left( \begin{array}{c}
                 M \\ m
                 \end{array} \right)^{2} } $, $M$ is the key pool size and $m$ is the key ring size of a sensor node. \\ 

\indent For the polynomial-pool based scheme~\cite{cf:04}, the probability of establishing a direct pairwise
key between two sensor nodes is
\begin{equation}
p_{poly-pool} = 1 - \frac{\left( \begin{array}{c}
                 s-s' \\ s'
                 \end{array} \right) }
        {\left( \begin{array}{c}
                 s \\ s'
                 \end{array} \right) } 
=1 - \prod_{i=0}^{s'-1} \frac{s-s'-i}{s-i}
\end{equation}
\noindent where $s$ is the polynomial-pool size and $s'$ is the number of shares given to a sensor node. Thus, we see that the EG and the $q$-composite schemes depend on $M$ and $m$. The polynomial-pool based scheme depends on $s$ and $s'$ and the maximum supported network size is bounded by $\frac{(t+1)s}{s'}$, where $t$ is the degree of the symmetric bivariate polynomial, whereas our scheme depends on the network size $n$ and the key ring size $m$.

\begin{figure}[h]
\centering
\includegraphics[scale=0.33,angle=270]{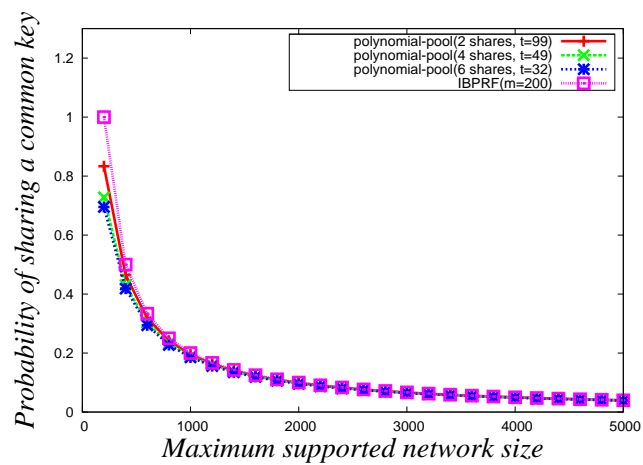}
\caption{The probability $p$ of establishing a common key v.s. the maximum supported network size $n$ in order to be resilient against node compromise. Assume that each sensor node is capable of holding $200$ keys.} \label{Figure: }
\end{figure}

\indent For comparison of network connectivity, we only consider the polynomial-pool based scheme because it is more resilient against node compromise than the EG scheme and the $q$-composite scheme. However, both the EG scheme and the $q$-composite scheme support networks of arbitarily big sizes. The relationship between the probability of establishing direct keys and the maximum supported network size for the polynomial-pool based scheme and our basic scheme (IBPRF) is shown in Figure 3. We assume that each sensor is capable of storing $200$ keys in its key ring. From this figure, it is very clear that our scheme provides better connectivity than the polynomial-pool based scheme in order to be resilient against node compromise.

\section{Conclusion}
Our basic scheme (IBPRF) is an alternative to direct key establishment of the bootstrapping protocol. Both IBPRF and the improved scheme guarantee that they have better trade-off between communication overhead, network connectivity and also resilience against node captures compared to the EG, the $q$-composite, and the polynomial-pool based schemes. Both schemes can also be adapted for mobile sensor networks by initiating direct key establishment phase and one can achieve reasonable connectivity by applying these schemes.

\bibliographystyle{plain}
\bibliography{sensor}

\end{document}